\documentclass[letterpaper,10pt,conference]{ieeeconf}
\IEEEoverridecommandlockouts
\usepackage{amsmath,amssymb,mathtools}
\usepackage{bm}
\usepackage{graphicx}
\usepackage[nameinlink,noabbrev]{cleveref}
\usepackage{algorithm}
\usepackage{algpseudocode}
\usepackage{booktabs}
\usepackage{microtype}
\usepackage{xcolor}
\usepackage{cite}
\usepackage{tikz}
\usepackage{mathrsfs}
\usetikzlibrary{arrows.meta,positioning,fit}
\usepackage[american]{circuitikz}

\usepackage{balance}
\usepackage{amsmath}
\allowdisplaybreaks

\usepackage[caption=false]{subfig}
\usepackage[font=normalsize]{caption}
\def\p(#1|#2){p(#1\,|\,#2)}

\begin{document}

\title{Recursive Parameter Identification of Nonlinear Stochastic State-Space Models via Sequentialized Ensemble Kalman Inversion}

\author{
    Farzaneh Barat\textsuperscript{1}, Rachel Carter\textsuperscript{1}, Sara Wilson\textsuperscript{1}, and Huazhen Fang\textsuperscript{2}%
    \thanks{\textsuperscript{1}F. Barat, R. Carter, and S. Wilson are with the Department of Mechanical Engineering, University of Kansas, Lawrence, KS 66045, USA. {\tt\small \{barat, rachel.carter, sewilson\}@ku.edu}}%
    \thanks{\textsuperscript{2}H. Fang is with the Department of Mechanical Engineering, Michigan State University, East Lansing, MI 48824, USA. {\tt\small hfang@msu.edu}}%
}

\maketitle

\begin{abstract}
Recursive parameter identification in nonlinear stochastic state-space models is challenging because unknown parameters affect the measurements through latent-state dynamics. This paper develops a sequentialized ensemble Kalman inversion method for recursive parameter identification from streaming measurements. The proposed method represents the parameter posterior by an evolving ensemble and updates it sequentially as new measurements become available. For each parameter ensemble member, implicit particle filtering is used to approximate the predictive observation statistics required for parameter correction. This formulation enables recursive parameter learning. The method is evaluated on a strongly nonlinear benchmark, with comparison to several existing methods, and then on a nonlinear double-capacitor lithium-ion battery model. The numerical results demonstrate accurate recursive parameter identification and latent-state estimation.
\end{abstract}

\section{Introduction}

State-space models (SSMs) provide a powerful framework for representing dynamic systems across a wide range of engineering and scientific domains. Their practical application, however, requires the identification of unknown model parameters from measurement data. This problem, known as parameter estimation, is challenging due to its nonlinear nature and the coupling between latent system states and unknown parameters~\cite{Ljung:1999}. Despite sustained research interest over several decades, it remains challenging~\cite{COURTS:2023}.

In the literature, the majority of research has focused on iterative identification for SSMs, wherein model parameters are successively refined using the entire dataset. Prediction-error methods formulate a model prediction-error minimization problem and use gradient-based optimization to estimate the unknown parameters~\cite{Doucet:2003,Ionides:2006}. In nonlinear stochastic SSMs, however, gradient evaluation can become computationally demanding because the likelihood depends on unobserved states. Maximum-likelihood (ML) methods therefore often rely on particle filtering or related inference techniques to approximate the required likelihood information~\cite{Doucet:2003,Ionides:2006,Poyiadjis:2011}. Alternatively, expectation--maximization methods~\cite{SCHON:2011,Picchini:2018} avoid direct likelihood-gradient evaluation, while maximum a posteriori (MAP) estimation incorporates prior information~\cite{Andrieu:2010,Kantas:2015}. Black-box optimization approaches, including Bayesian optimization, have also been investigated~\cite{tu:2024}. These methods are generally iterative and process a batch of measurements through multiple optimization iterations, making them less directly suited to recursive parameter updating from streaming data.

Another important line of research is recursive identification, in which parameter estimates are updated sequentially as new measurements become available. Such capability is important in applications including condition monitoring, state estimation, and fault detection, where streaming measurements must be processed online. A common approach is to augment the system state with the unknown parameters and apply a nonlinear Kalman filter or particle filter to estimate the joint state-parameter vector~\cite{Carrassi:2011,Naets:2015}. For static parameters, augmented-state approaches commonly employ a constant or random-walk parameter model; when artificial parameter process noise is introduced to maintain adaptation, its tuning can significantly influence the resulting estimates. Another class of approaches uses dual estimation, in which separate state and parameter estimators interact recursively~\cite{MORADKHANI:2005,Santitissadeekorn:2014,hess:2016}. Unlike augmented-state approaches, dual estimators do not necessarily require artificial parameter dynamics. Nevertheless, their coupled state and parameter updates often require cumbersome tuning, and errors in one estimator may propagate into the other.

In this paper, we extend our prior work in~\cite{barat:2026} to use ensemble Kalman inversion (EnKI) to recursively estimate the unknown parameters in a nonlinear stochastic SSM. EnKI leverages an ensemble of samples to represent the posterior parameter distribution while updating it iteratively. In~\cite{barat:2026}, we have developed an iterative EnKI-based SSM identification method. In a departure, we propose a sequentialized EnKI (SEnKI) method to enable recursive identification. By design, the SEnKI method performs the parameter update sequentially by assimilating the measurement data online.

The proposed method maintains an ensemble representation of the parameter distribution and recursively updates the ensemble as new measurements become available. For each parameter ensemble member, particle-based state inference combined with local unscented transforms is used to approximate the predictive observation statistics required for the parameter correction. This structure avoids explicit state--parameter augmentation. The method is evaluated first on a nonlinear benchmark against augmented and dual extended/unscented Kalman filters estimators and is then demonstrated on a nonlinear double-capacitor (NDC) lithium-ion battery model.

The remainder of the paper is organized as follows. Section~\ref{sec:setup} formulates the problem. Section~\ref{sec:senki} develops SEnKI. Sections~\ref{sec:ungm} and~\ref{sec:ndc} present the numerical studies, including a nonlinear benchmark comparison and an application to lithium-ion battery. Section~\ref{sec:conclusion} concludes the paper.


\section{Problem Formulation}
\label{sec:setup}

Consider the nonlinear stochastic SSM
\begin{subequations}
\label{eq:ssm_model}
\begin{align}
\bm{x}_{k+1}
&=
f(\bm{x}_k,\bm{u}_k,\bm{\theta})+\bm{w}_k,
\label{eq:ssm_state}\\
\bm{y}_k
&=
h(\bm{x}_k,\bm{u}_k,\bm{\theta})+\bm{v}_k,
\label{eq:ssm_obs}
\end{align}
\end{subequations}
where $\bm{x}_k\in\mathbb{R}^{n}$ is the state, $\bm{u}_k$ is the input,
$\bm{y}_k\in\mathbb{R}^{n_y}$ is the measurement, $\bm{\theta}\in\mathbb{R}^{s}$ is the unknown parameter vector, and $\bm{w}_k\sim\mathcal N(\bm{0},\bm Q)$ and $\bm{v}_k\sim\mathcal N(\bm{0},\bm R)$ are independent process and measurement noises, respectively. Known inputs are omitted from the conditioning notation for brevity.

For the SSM in~\eqref{eq:ssm_model}, the objective is to estimate
$\bm{\theta}$ recursively from the sequential measurements
\begin{align*}
\bm{y}_{1:k}:=\{\bm{y}_1,\dots,\bm{y}_k\}.
\end{align*}
The estimate is given by
\begin{align*}
\hat{\bm{\theta}}_k
=
\mathbb{E}\!\left[\bm{\theta}\mid \bm{y}_{1:k}\right] 
=
\int \bm{\theta}\,\p(\bm{\theta} | \bm{y}_{1:k})\,d\bm{\theta},
\end{align*}
which minimizes the mean-square estimation error~\cite{AndersonMoore:2013}. By Bayes' rule, $\p(\bm{\theta}|\bm{y}_{1:k})$ evolves recursively as
\begin{align} 
\label{Bayes-update}
\p(\bm{\theta} | \bm{y}_{1:k})
\propto
\p(\bm{y}_k | \bm{\theta},\bm{y}_{1:k-1})\,
\p(\bm{\theta} | \bm{y}_{1:k-1}).
\end{align}

For nonlinear SSMs, \eqref{Bayes-update} is generally intractable. To obtain a tractable parameter update, we introduce the following approximation, inspired by EnKI~\cite{barat:2026}:
\begin{align}
\begin{bmatrix}
\bm{\theta}\\
\bm{y}_k
\end{bmatrix}
\Bigg|\bm{y}_{1:k-1}
\approx
\mathcal{N}
\left(
\begin{bmatrix}
\hat{\bm{\theta}}_{k-1}\\
\hat{\bm{y}}_{k|k-1}
\end{bmatrix},
\begin{bmatrix}
\bm{C}_{k-1}^{\bm{\theta}\bm{\theta}} & \bm{C}_{k|k-1}^{\bm \theta \bm y}\\
\left(\bm{C}_{k|k-1}^{ \bm{\theta} \bm y}\right)^\top & \bm{C}_{k|k-1}^{\bm y\bm y}
\end{bmatrix}
\right),
\label{eq:joint_gaussian_theta_y}
\end{align}
where
\begin{align*}
\hat{\bm{\theta}}_{k-1}
&=
\mathbb{E}\!\left[\bm{\theta} \,|\, \bm{y}_{1:k-1}\right]
=
\int \bm{\theta}\, \p(\bm{\theta} | \bm{y}_{1:k-1})
\,d\bm{\theta}, \\
\hat{\bm{y}}_{k|k-1}
&=
\mathbb{E}\!\left[\bm{y}_k \,|\, \bm{y}_{1:k-1}\right]
=
\int
\bm{y}_k\, \p(\bm{y}_k | \bm{y}_{1:k-1}) \,d\bm{y}_k, \\
\bm{C}_{k-1}^{\bm{\theta}\bm{\theta}} &=
\operatorname{Cov}\!\left(\bm{\theta} \,|\, \bm{y}_{1:k-1}\right)
\nonumber\\
&= \int
(\bm{\theta}-\hat{\bm{\theta}}_{k-1})
(\bm{\theta}-\hat{\bm{\theta}}_{k-1})^\top
\p(\bm{\theta} | \bm{y}_{1:k-1})
\,d\bm{\theta}, \\
\bm{C}_{k|k-1}^{\bm y\bm y} &= \operatorname{Cov}\!\left(\bm{y}_k \,|\, \bm{y}_{1:k-1}\right)
\nonumber\\
&= \int (\bm{y}_k-\hat{\bm{y}}_{k|k-1})
(\bm{y}_k-\hat{\bm{y}}_{k|k-1})^\top \\
&\qquad\qquad\qquad\qquad\qquad\qquad \times
\p(\bm{y}_k | \bm{y}_{1:k-1})
\,d\bm{y}_k, \\
\bm{C}_{k|k-1}^{\bm{\theta} \bm y}
&= \operatorname{Cov}\!\left(\bm{\theta},\bm{y}_k|\bm{y}_{1:k-1}\right)
\nonumber\\
&= \iint (\bm{\theta}-\hat{\bm{\theta}}_{k-1})
(\bm{y}_k-\hat{\bm{y}}_{k|k-1})^\top \\
&\qquad\qquad\qquad\qquad\quad \times
\p(\bm{\theta},\bm{y}_k | \bm{y}_{1:k-1})
\,d\bm{\theta}\,d\bm{y}_k.
\end{align*}
Conditioning~\eqref{eq:joint_gaussian_theta_y} on $\bm y_k$ gives
\begin{align}
\hat{\bm{\theta}}_k
&=
\hat{\bm{\theta}}_{k-1}
+
\bm{C}_{k|k-1}^{\bm{\theta}\bm y}
\left(
\bm{C}_{k|k-1}^{\bm y\bm y}
\right)^{-1}
\left(
\bm y_k-\hat{\bm y}_{k|k-1}
\right).
\label{eq:gaussian_parameter_mean_update}
\end{align}
Equation~\eqref{eq:gaussian_parameter_mean_update} requires
$\hat{\bm y}_{k|k-1}$,
$\bm C_{k|k-1}^{\bm y\bm y}$, and
$\bm C_{k|k-1}^{\bm\theta\bm y}$, which are generally unavailable in closed form for nonlinear SSMs with latent states.


\section{SEnKI-Based Recursive Parameter Estimation}
\label{sec:senki}

In this section, we first introduce the ensemble-based recursive parameter update procedure. Along with the parameter update, the latent states must also be estimated. To perform the state estimation,  we leverage an implicit particle filter proposed in~\cite{ASKARI:2022:A}.

\subsection{Ensemble-Based Parameter Update}

Suppose that $\p(\bm{\theta}|\bm{y}_{1:k-1})$ is represented by
$\{\bm{\theta}_{k-1}^{j}\}_{j=1}^{N_s}$. For each
$\bm{\theta}_{k-1}^{j}$, define
\begin{align}
\hat{\bm{y}}_{k|k-1}^{j}
=
\mathbb{E}\!\left[
\bm{y}_k
\mid
\bm{\theta}_{k-1}^{j},
\bm{y}_{1:k-1}
\right].
\label{eq:conditional_output_mean}
\end{align}
Based on~\eqref{eq:gaussian_parameter_mean_update},  the SEnKI parameter update  is given by
\begin{align}
\bm{\theta}_{k}^{j}
=
\bm{\theta}_{k-1}^{j}
+
\bm{C}_{k|k-1}^{\bm{\theta}\bm{y}}
\left(
\bm{C}_{k|k-1}^{\bm{y}\bm{y}}
\right)^{-1}
\left(
\bm{y}_k-\hat{\bm{y}}_{k|k-1}^{j}
\right),
\label{eq:senki_parameter_update}
\end{align}
where
\begin{subequations}
\label{eq:cov}
\begin{align}
\bar{\bm{\theta}}_{k-1}
&=
\frac{1}{N_s}
\sum_{j=1}^{N_s}
\bm{\theta}_{k-1}^{j},
\quad
\bar{\bm{y}}_{k|k-1}
=
\frac{1}{N_s}
\sum_{j=1}^{N_s}
\hat{\bm{y}}_{k|k-1}^{j},
\label{eq:ensemble_means}
\\
\bm{C}_{k|k-1}^{\bm{\theta}\bm{y}}
&=
\frac{1}{N_s-1}
\sum_{j=1}^{N_s}
\left(
\bm{\theta}_{k-1}^{j}
-
\bar{\bm{\theta}}_{k-1}
\right) \nonumber\\
&\qquad\qquad\qquad\qquad\times
\left(
\hat{\bm{y}}_{k|k-1}^{j}
-
\bar{\bm{y}}_{k|k-1}
\right)^\top.
\label{eq:ensemble_cross_covariance}
\\
\bm{C}_{k|k-1}^{\bm y\bm y}
&=
\frac{1}{N_s-1}
\sum_{j=1}^{N_s}
\left(
\hat{\bm{y}}_{k|k-1}^{j}
-
\bar{\bm{y}}_{k|k-1}
\right)
\nonumber\\
&\qquad\qquad\qquad\qquad\times
\left(
\hat{\bm{y}}_{k|k-1}^{j}
-
\bar{\bm{y}}_{k|k-1}
\right)^\top.
\label{eq:ensemble_output_covariance}
\end{align}
\end{subequations}

\subsection{Implicit Particle Filtering for State Estimation}

The parameter update in~\eqref{eq:senki_parameter_update}--\eqref{eq:cov} requires to compute $\hat{\bm{y}}_{k|k-1}^{j}$, the definition of which is shown in~\eqref{eq:conditional_output_mean}. 

For each $\bm{\theta}_{k-1}^{j}$, we have
\begin{align}
&\p(\bm{y}_k | \bm{\theta}_{k-1}^{j},
\bm{y}_{1:k-1})
\nonumber\\
&\quad=
\int
\p(\bm{y}_k | \bm{x}_k,\bm{\theta}_{k-1}^{j})
\p(\bm{x}_k | \bm{\theta}_{k-1}^{j},
\bm{y}_{1:k-1})
\,d\bm{x}_k,
\label{eq:predictive_obs_distribution}
\end{align}
where
\begin{align}
&\p(\bm{x}_k|\bm{\theta}_{k-1}^{j},
\bm{y}_{1:k-1})
\nonumber\\
&\quad=
\int
\p(\bm{x}_k|\bm{x}_{k-1},
\bm{\theta}_{k-1}^{j})
\p(\bm{x}_{k-1}|
\bm{\theta}_{k-1}^{j},
\bm{y}_{1:k-1})
\,d\bm{x}_{k-1}.
\label{eq:predictive_state_distribution}
\end{align}
For nonlinear SSMs, \eqref{eq:predictive_obs_distribution}--\eqref{eq:predictive_state_distribution} are generally unavailable in closed form. As such,  we perform the state filtering, drawing upon the implicit particle filtering method proposed in~\cite{ASKARI:2022:A}. Here, given each $\bm{\theta}_{k-1}^{j}$, we empirically represent $\p(\bm{x}_{k-1}|\bm{\theta}_{k-1}^{j},\bm{y}_{1:k-1})$ using a set of particles:
\begin{align}
&\p(\bm{x}_{k-1}|
\bm{\theta}_{k-1}^{j},\bm{y}_{1:k-1})
\approx
\sum_{i=1}^{N_p}
w_{k-1}^{i,j}
\delta\!\left(
\bm{x}_{k-1}-\bm{x}_{k-1}^{i,j}
\right),
\label{eq:state_particle_approx}
\end{align}
where $\bm{x}_{k-1}^{i,j}$ is also associated with a covariance $\bm{P}_{k-1}^{i,j}$. 
Then, the filtering procedure at time $k$ includes the following steps.

\emph{Step 1: State prediction.}
We propagate $\bm{x}_{k-1}^{i,j}$ forward using the unscented transform (UT) (see~\cite{ASKARI:2022:A}) to obtain a new set of particles:
\begin{align}
&\left[
\bm{m}_{\bm{x},k|k-1}^{i,j},
\bm{P}_{\bm{x},k|k-1}^{i,j}
\right]
\nonumber\\
&\quad=
\mathcal{UT}\!\left(
f(\cdot,\bm{u}_{k-1},\bm{\theta}_{k-1}^{j}),
\bm{x}_{k-1}^{i,j},
\bm{P}_{k-1}^{i,j},
\bm{Q}
\right).
\label{eq:ut_state_prediction}
\end{align}

\emph{Step 2: Output prediction.}
Using the UT transform, we generate, a new set of particles to approximate $\p(\bm{y}_k|\bm{\theta}_{k-1}^{j},\bm{y}_{1:k-1})$:
\begin{align}
&\left[
\bm{m}_{\bm{y},k|k-1}^{i,j},
\bm{P}_{\bm{x}\bm{y},k|k-1}^{i,j},
\bm{P}_{\bm{y},k|k-1}^{i,j}
\right]
\nonumber\\
&\quad=
\mathcal{UT}\!\left(
h(\cdot,\bm{u}_{k},\bm{\theta}_{k-1}^{j}),
\bm{m}_{\bm{x},k|k-1}^{i,j},
\bm{P}_{\bm{x},k|k-1}^{i,j},
\bm{R}
\right).
\label{eq:ut_obs_prediction}
\end{align}
Given the above, we can compute $\hat{\bm{y}}_{k|k-1}^{j}$ by
\begin{align}
\hat{\bm{y}}_{k|k-1}^{j}
&=
\sum_{i=1}^{N_p}
w_{k-1}^{i,j}
\bm{m}_{\bm{y},k|k-1}^{i,j}.
\label{eq:predictive_mean_particle}
\end{align}

\emph{Step 3: State update.}
We apply the Kalman update to $\bm{x}_{k-1}^{i,j}$ as below to obtain $\bm{x}_{k}^{i,j}$, which approximately represents
$\p(\bm{x}_k|\bm{\theta}_{k-1}^{j},\bm{y}_{1:k})$:
\begin{subequations}
\label{eq:local_update}
\begin{align}
\bm{K}_{\bm{x},k}^{i,j}
&=
\bm{P}_{\bm{x}\bm{y},k|k-1}^{i,j}
\left(
\bm{P}_{\bm{y},k|k-1}^{i,j}
\right)^{-1},
\label{eq:local_state_gain}\\
\bm{m}_{\bm{x},k|k}^{i,j}
&=
\bm{m}_{\bm{x},k|k-1}^{i,j}
+
\bm{K}_{\bm{x},k}^{i,j}
\left(
\bm{y}_k-
\bm{m}_{\bm{y},k|k-1}^{i,j}
\right),
\label{eq:local_state_update_mean}\\
\bm{P}_{\bm{x},k|k}^{i,j}
&=
\bm{P}_{\bm{x},k|k-1}^{i,j}
-
\bm{K}_{\bm{x},k}^{i,j}
\bm{P}_{\bm{y},k|k-1}^{i,j}
\left(\bm{K}_{\bm{x},k}^{i,j}\right)^\top.
\label{eq:local_state_update_cov}
\end{align}
\end{subequations}
Following the implicit sampling formulation in~\cite{ASKARI:2022:A}, the updated
state particle is generated as
\begin{align}
\bm{x}_k^{i,j}
&=
\bm{m}_{\bm{x},k|k}^{i,j}
+
\left(
\bm{P}_{\bm{x},k|k}^{i,j}
\right)^{1/2}
\bm{\xi}_k^{i,j},
\label{eq:uipf_particle}
\end{align}
where
\begin{align*}
\bm{\xi}_k^{i,j}
\sim
\mathcal{N}(\bm{0},\gamma\bm{I}),
\qquad
0<\gamma\ll1,
\end{align*}
so that the generated particles are concentrated in high-probability
regions. The corresponding normalized weight is
\begin{align}
w_k^{i,j}
&=
\frac{
w_{k-1}^{i,j}
\mathcal{N}\!\left(
\bm{y}_k;
\bm{m}_{\bm{y},k|k-1}^{i,j},
\bm{P}_{\bm{y},k|k-1}^{i,j}
\right)
}{
\displaystyle
\sum_{\ell=1}^{N_p}
w_{k-1}^{\ell,j}
\mathcal{N}\!\left(
\bm{y}_k;
\bm{m}_{\bm{y},k|k-1}^{\ell,j},
\bm{P}_{\bm{y},k|k-1}^{\ell,j}
\right)
}.
\label{eq:uipf_weight}
\end{align}
The particles are resampled when the effective sample size falls below
the prescribed threshold.

Following the above filtering procedure, the parameter update is performed using
\eqref{eq:senki_parameter_update}. The current particle set
$\{\bm{x}_k^{i,j}\}_{i=1}^{N_p}$ is then used to approximately represent
$\p(\bm{x}_k|\bm{\theta}_{k}^{j},\bm{y}_{1:k})$ for the next recursion.

The parameter and state estimates are then given by
\begin{subequations}
\begin{align}
\hat{\bm{\theta}}_k
&\coloneqq
\bar{\bm{\theta}}_k
=
\frac{1}{N_s}
\sum_{j=1}^{N_s}
\bm{\theta}_k^{j},
\label{eq:parameter_estimate}\\
\hat{\bm{x}}_k
&=
\frac{1}{N_s}
\sum_{j=1}^{N_s}
\sum_{i=1}^{N_p}
w_k^{i,j}
\bm{x}_k^{i,j}.
\label{eq:state_estimate}
\end{align}
\end{subequations}

The complete SEnKI procedure, including U-IPF-based state filtering and recursive parameter correction, is summarized in
Algorithm~\ref{alg:seq_enki_uipf}.

\begin{algorithm}[h]
\caption{SEnKI for Recursive Parameter Identification}
\label{alg:seq_enki_uipf}
\small
\begin{algorithmic}[1]

\Require $\{\bm{y}_k,\bm{u}_k\}_{k=1}^{N}$, $N_s$, $N_p$,
$\{\bm{\theta}_0^{j}\}_{j=1}^{N_s}$
\Require $\{\bm{x}_0^{i,j},\bm{P}_0^{i,j},w_0^{i,j}\}$
\For{$k=1$ to $N$}
    \For{$j=1$ to $N_s$}
        \For{$i=1$ to $N_p$}
            \State Compute state and output predictions using \eqref{eq:ut_state_prediction}--\eqref{eq:ut_obs_prediction}
        \EndFor
        \State Compute
        $\hat{\bm{y}}_{k|k-1}^{j}$ using
        \eqref{eq:predictive_mean_particle}
        \For{$i=1$ to $N_p$}
            \State Compute the state update using
            \eqref{eq:local_update}
            \State Generate $\bm{x}_k^{i,j}$ using
            \eqref{eq:uipf_particle}
        \EndFor
        \State Update $\{w_k^{i,j}\}_{i=1}^{N_p}$ using
        \eqref{eq:uipf_weight}

    \EndFor
    \State Compute the ensemble moments using
    \eqref{eq:ensemble_means},
    \eqref{eq:ensemble_cross_covariance}, and
    \eqref{eq:ensemble_output_covariance}
    \For{$j=1$ to $N_s$}
        \State Update $\bm{\theta}_k^{j}$ using
        \eqref{eq:senki_parameter_update}
    \EndFor
    \State Compute $\hat{\bm{\theta}}_k$ and $\hat{\bm{x}}_k$ using
    \eqref{eq:parameter_estimate}--\eqref{eq:state_estimate}
    \For{$j=1$ to $N_s$}
        \State Resample the state particles if needed
    \EndFor
\EndFor
\end{algorithmic}
\end{algorithm}

\section{Nonlinear Benchmark and Comparative Evaluation}
\label{sec:ungm}

To evaluate the proposed method on a strongly nonlinear SSM, we modify the classical univariate nonstationary growth model (UNGM)~\cite{Kitagawa:1987}.
Specifically, the four coefficients in the standard state equation are treated as unknown parameters
$\theta_1,\ldots,\theta_4$, and an additional unknown parameter
$\theta_5$ is introduced into the measurement equation so that unknown parameters appear in both the state and measurement models. The resulting model is
\begin{align*}
x_k &=
\theta_1 x_{k-1}
+\theta_2\frac{x_{k-1}}{1+x_{k-1}^2}
+\theta_3\cos(\theta_4 k)+w_k,
\\
y_k &=
\frac{x_k^2}{20}+\theta_5 x_k+v_k,
\end{align*}
where $w_k\sim\mathcal{N}(0,Q)$ and $v_k\sim\mathcal{N}(0,R)$.

For comparison, we consider two augmented state--parameter estimators, the augmented extended Kalman filter (AEKF) and augmented unscented Kalman filter (AUKF)~\cite{Carrassi:2011,Naets:2015}, and two dual estimators, the dual extended Kalman filter (DEKF) and dual unscented Kalman filter (DUKF)~\cite{MORADKHANI:2005,Santitissadeekorn:2014,hess:2016}.

For a consistent comparison, all methods use the same initial state, parameter prior, noise settings, and admissible parameter bounds. The simulation uses $N=1000$, $Q=10$, and $R=1$. The prior mean is randomly perturbed by $30\%$ from the true parameters, with initial standard deviations equal to $20\%$ of the corresponding true values. SEnKI uses $N_s=400$ parameter samples and $N_p=300$ state particles. Performance is evaluated using the mean relative parameter error and
the state root-mean-square error (RMSE).

\begin{figure}[htbp]
\centering
\includegraphics[width=\linewidth]{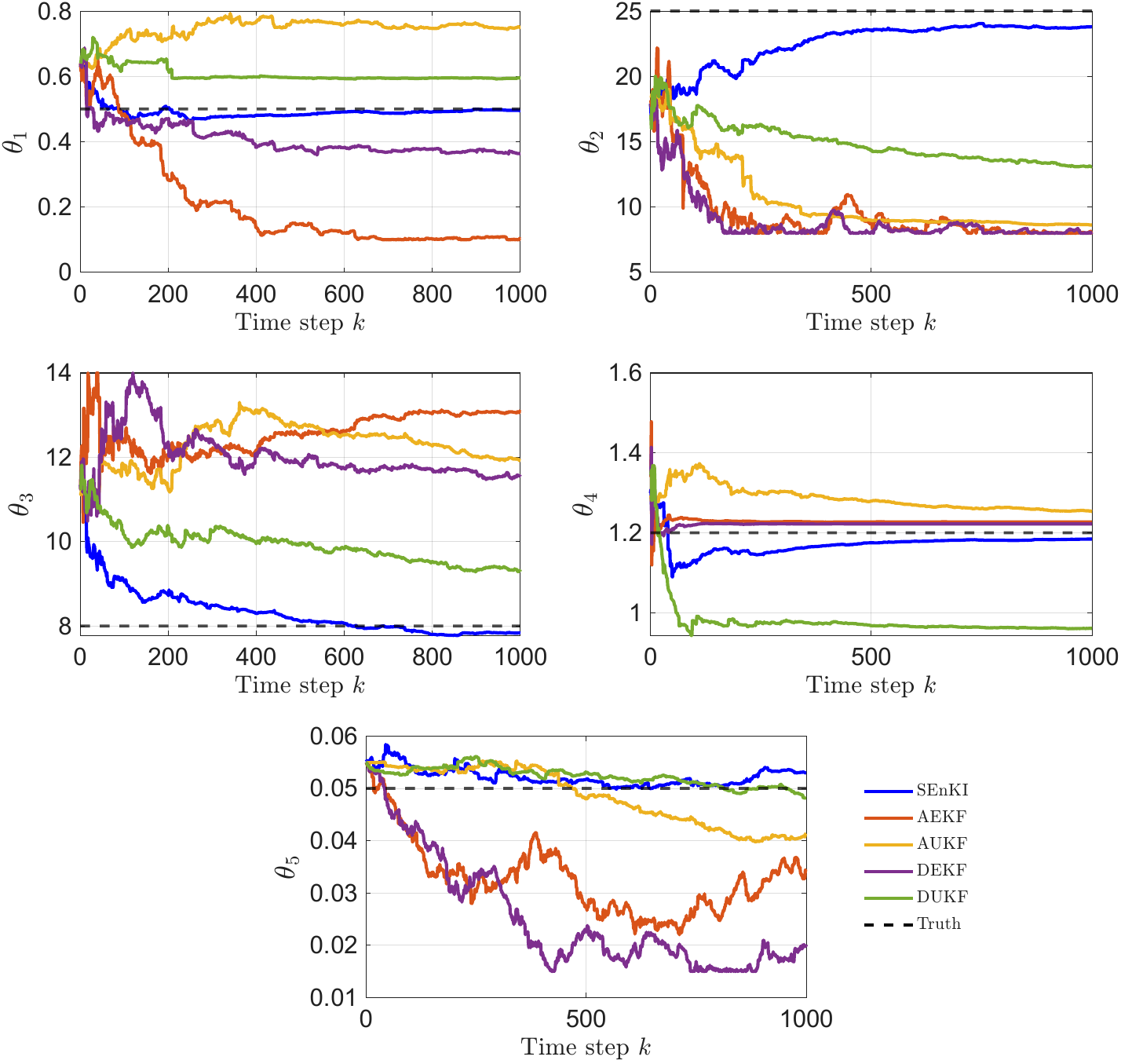}
\caption{Recursive estimates of the UNGM parameters obtained using SEnKI, AEKF, AUKF, DEKF, and DUKF.}
\label{fig:ungm_comparison}
\end{figure}

Figure~\ref{fig:ungm_comparison} shows the recursive parameter estimates. Table~\ref{tab:ungm_comparison}(a) reports the final mean relative parameter error and state RMSE, while Table~\ref{tab:ungm_comparison}(b) compares the final parameter estimates with their true values.

\begin{table}[htbp]
\centering
\caption{Quantitative comparison on the nonlinear UNGM benchmark.}
\label{tab:ungm_comparison}
\renewcommand{\arraystretch}{1.08}
\setlength{\tabcolsep}{6pt}

\begin{tabular}{lcc}
\hline
\multicolumn{3}{c}{\textbf{(a) Overall estimation performance}}\\
\hline
Method & Mean relative error (\%) & State RMSE\\
\hline
SEnKI & \textbf{3.00} & \textbf{11.50}\\
AEKF     & 48.99 & 15.78\\
AUKF     & 37.21 & 13.03\\
DEKF     & 40.15 & 16.25\\
DUKF     & 21.28 & 13.69\\
\hline
\end{tabular}

\vspace{1.5mm}

\begin{tabular}{c|c|ccccc}
\hline
\multicolumn{7}{c}{\textbf{(b) Final parameter estimates}}\\
\hline
Parameter & True & SEnKI & AEKF & AUKF & DEKF & DUKF\\
\hline
$\theta_1$ & 0.500 & \textbf{0.495} & 0.103 & 0.750 & 0.364 & 0.595\\
$\theta_2$ & 25.000 & \textbf{23.793} & 8.051 & 8.631 & 8.083 & 13.102\\
$\theta_3$ & 8.000 & \textbf{7.839} & 13.085 & 11.917 & 11.543 & 9.290\\
$\theta_4$ & 1.200 & \textbf{1.184} & 1.227 & 1.254 & 1.222 & 0.960\\
$\theta_5$ & 0.050 & 0.0530 & 0.0340 & 0.0414 & 0.0201 & \textbf{0.0481}\\
\hline
\end{tabular}
\end{table}

As shown in Table~\ref{tab:ungm_comparison}(a), SEnKI achieves a final mean relative parameter error of $3.00\%$ and a state RMSE of $11.50$, compared with parameter errors of $48.99\%$, $37.21\%$, $40.15\%$, and $21.28\%$ for AEKF,
AUKF, DEKF, and DUKF, respectively.
Table~\ref{tab:ungm_comparison}(b) further shows that the SEnKI estimates are closest to the true values for $\theta_1$--$\theta_4$, whereas DUKF gives the closest estimate of $\theta_5$. Under the considered simulation setting, these results show
more accurate joint state and parameter estimation with SEnKI.

\section{NDC Battery Model and Simulation Results}
\label{sec:ndc}

This section applies SEnKI to parameter estimation of the NDC battery model in~\cite{Tian:2021:IEEE}. The method is evaluated in simulation.

\subsection{NDC Battery Model}

\begin{figure}[htbp]
\centering
\ctikzset{ bipoles/thickness=1, bipoles/length=1.1cm }
\begin{circuitikz}[line width=0.7pt, scale=0.9, transform shape]

\draw (0,0) -- (3.25,0);
\draw (0,0) to[C,l=$C_b$] (0,2);
\draw (0,2) node[above] {$V_b$}
      to[R,l=$R_b$] (2,2) node[above] {$V_s$};
\draw (2,2) to[C,l_=$C_s$] (2,0);
\draw (2,2) -- (3.25,2) node[circ]{};
\draw[->] (3.0,2.4) -- (2.4,2.4);
\node at (2.8,2.6) {$I$};
\draw (2.87,-0.09) -- (3.23,-0.09);
\draw (2.92,-0.15) -- (3.18,-0.15);
\draw (2.97,-0.21) -- (3.13,-0.21);

\begin{scope}[xshift=4.7cm]
\draw (0,0) to[V,l=$\mathrm{OCV}$, invert] (0,2);
\draw (0,2) to[R,l=$R_o$] (2.5,2);
\draw (2.5,2) -- (3,2) node[circ]{};
\draw (0,0) -- (3,0) node[circ]{};
\draw[->] (3.0,2.4) -- (2.4,2.4);
\node at (2.7,2.6) {$I$};
\draw[<->] (3,0.2) -- (3,1.8);
\node at (3.3,1.0) {$V$};
\end{scope}
\end{circuitikz}

\caption{The NDC equivalent-circuit model.}
\label{fig:NDC}
\end{figure}
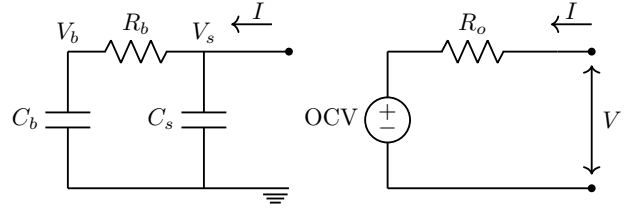

The NDC model is an equivalent-circuit model proposed in~\cite{Tian:2021:IEEE} to describe the electrical behavior and charge-storage dynamics of lithium-ion batteries. Figure~\ref{fig:NDC} shows its equivalent-circuit structure. The model includes two capacitors, $C_b$ and $C_s$, representing bulk and surface charge storage, respectively, connected through resistance $R_b$. The terminal voltage is determined by the open-circuit voltage associated with the surface capacitor and the ohmic voltage drop across the resistance $R_o$.

The continuous-time dynamics of the NDC model are given by
\begin{align*}
\dot V_b(t)
&=
-\frac{1}{C_b R_b}V_b(t)
+\frac{1}{C_b R_b}V_s(t),\\
\dot V_s(t)
&=
\frac{1}{C_s R_b}V_b(t)
-\frac{1}{C_s R_b}V_s(t)
+\frac{1}{C_s}I(t),
\end{align*}
and the terminal voltage is expressed as
\begin{align*}
V(t)=h_{\mathrm{OCV}}(V_s(t))+R_o I(t),
\end{align*}
where $V_b(t)$ and $V_s(t)$ are the voltages across $C_b$ and $C_s$, respectively, and $h_{\mathrm{OCV}}(\cdot)$ denotes the nonlinear open-circuit-voltage map, represented by a fifth-order polynomial with known coefficients~\cite{Tian:2021:IEEE}. Here, $I(t)$ is the applied current, with $I(t)<0$ for discharging and $I(t)>0$ for charging.

Define the state, input, output, and parameter vector as
\begin{align*}
\bm{x}(t)
&=
\begin{bmatrix}
V_b(t)\\
V_s(t)
\end{bmatrix},
\qquad
u(t)=I(t),
\qquad
y(t)=V(t),\\
\bm{\theta}
&=
\begin{bmatrix}
C_b & C_s & R_b & R_o
\end{bmatrix}^{\!\top},
\end{align*}
The continuous-time NDC model is discretized with sampling interval
$\Delta t=1~\mathrm{s}$.

\subsection{Simulation Results}
\label{sec:simulation}

The proposed method is evaluated using synthetic data generated from the NDC model under a dynamic current profile. The nominal NDC model parameters reported in~\cite{Tian:2021:IEEE}, identified from a Nickel Cobalt Aluminum Oxide (NCA) lithium-ion cell, are used as the true parameter values in the simulation study. The cell has a rated capacity of $3.3~\mathrm{Ah}$ and an operating voltage range of $4.2$ V to $2.5$ V. The input current is sampled at $\Delta t = 1$ s and scaled to the range $-4$ A to $1$ A.

Further, Gaussian process and measurement noises are added with covariances
$\bm{Q}=\operatorname{diag}(10^{-7},10^{-7})$ and $R=10^{-4}$,
respectively. The initial true state is $\bm{x}_0=[1,\,1]^\top$. The parameter prior mean is generated as
\begin{align*}
\bm{\mu}_0
=
\bm{\theta}_{\mathrm{true}}
+
0.3\,\operatorname{diag}(\bm{\theta}_{\mathrm{true}})\bm{\varepsilon},
\qquad
\bm{\varepsilon}\sim\mathcal{N}(\bm{0},\bm{I}),
\end{align*}
and the initial parameter ensemble is sampled from $\mathcal{N}(\bm{\mu}_0,\bm{\Sigma}_0)$, where $\bm{\Sigma}_0=
\operatorname{diag}\!\left(0.2\,\bm{\theta}_{\mathrm{true}}\right)^2$.
The initial state particles are sampled from
$\mathcal{N}(\bm{x}_0,\bm{P}_0)$ with
$\bm{P}_0=\operatorname{diag}(10^{-3},10^{-3})$, and $\bm{P}_0^{i,j}=\bm{P}_0$ for all $(i,j)$. SEnKI uses $N_s=100$ parameter samples and $N_p=200$ state particles.
Resampling is performed when the effective sample size falls below $0.5N_p$.
The UT parameters are $\alpha=0.6$, $\beta=2.0$, and $\kappa=0$.
After each parameter update, the samples are projected onto physically
admissible bounds.

Table~\ref{tab:param_results} presents the final parameter estimates alongside the true values used for synthetic data generation. The identified parameters are all close to the true values, with relative errors below $2\%$ for all four parameters. These results are consistent with the NDC model structure: $R_o$ appears directly in the terminal-voltage equation and is more identifiable from voltage measurements, whereas $C_b$, $C_s$, and $R_b$ are less identifiable because they influence the voltage indirectly through the latent-state dynamics.

\begin{table}[htbp]
\centering
\caption{Comparison between the true and identified parameters for the NDC model.}
\label{tab:param_results}
\begin{tabular}{lccc}
\toprule
\textbf{Parameter} & \textbf{True value} & \textbf{Identified value} & \textbf{Relative Error (\%)} \\
\midrule
$C_b$ [F] & 9960  & 9772   & 1.888 \\
$C_s$ [F] & 1154   & 1160   & 0.520 \\
$R_b$ [$\Omega$] & 0.0366  & 0.03677 & 0.464 \\
$R_o$ [$\Omega$] & 0.1130 & 0.11308 & 0.071 \\
\bottomrule
\end{tabular}
\end{table}

\begin{figure}[htbp]
\vspace{2mm}
\centering
\includegraphics[width=\linewidth]{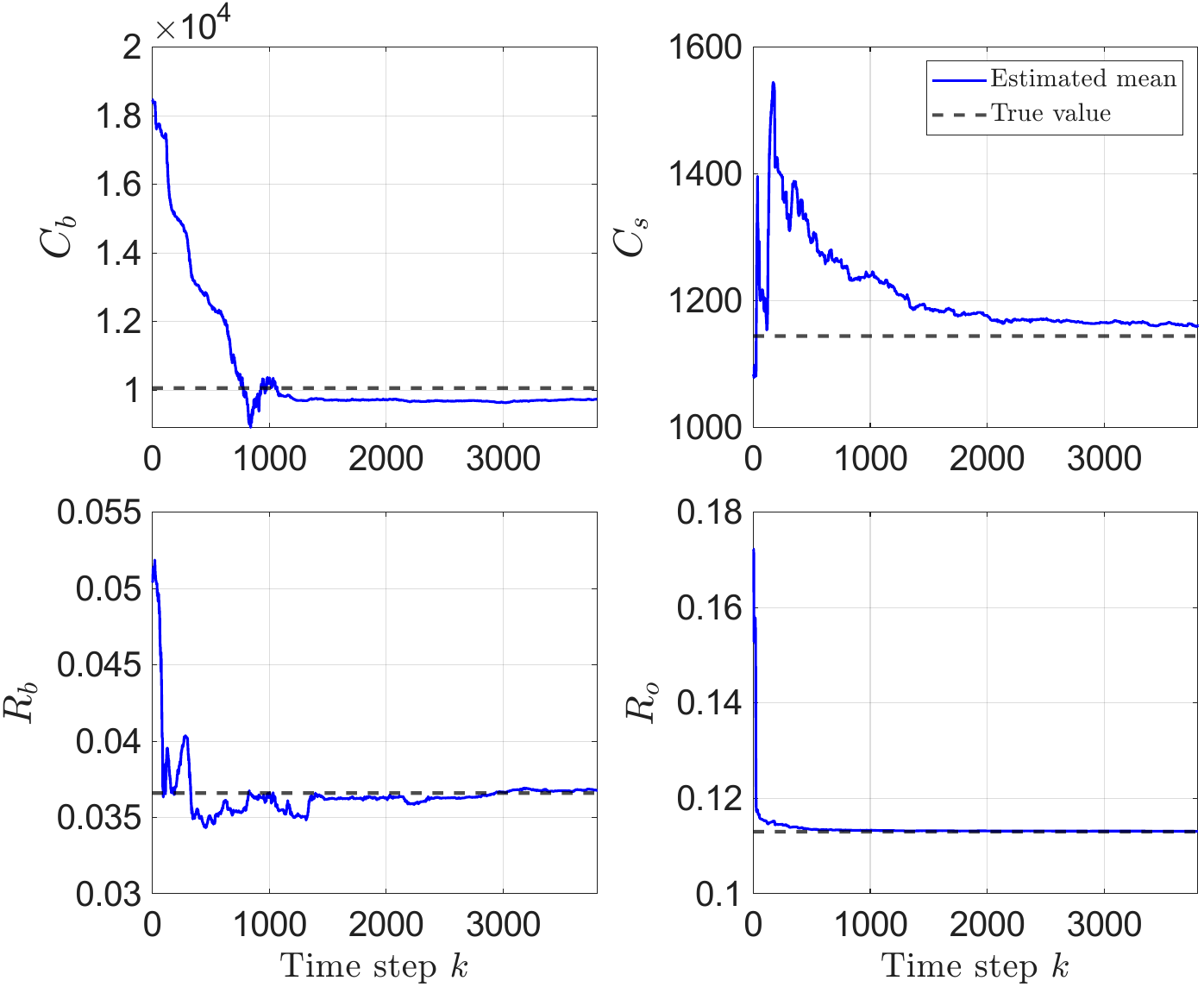}
\caption{Sequential evolution of the ensemble mean for the four estimated NDC parameters, showing transient adaptation and convergence toward the true values.}
\label{fig:param_traces}
\end{figure}

Figure~\ref{fig:param_traces} shows the sequential evolution of the ensemble mean for each parameter. Starting from an initially mismatched prior, the estimates first undergo a transient adaptation phase and then move toward the true values. The convergence speed differs among parameters: $R_o$ approaches its true value more directly because it influences the measured voltage more explicitly, whereas $C_b$, $C_s$, and $R_b$ are identified more indirectly through the latent-state dynamics. After the transient phase, the parameter trajectories remain close to their final values, showing that the proposed recursive update can refine the parameter ensemble as new measurements are assimilated.

\begin{figure}[htbp]
\centering
\includegraphics[width=0.95\linewidth]{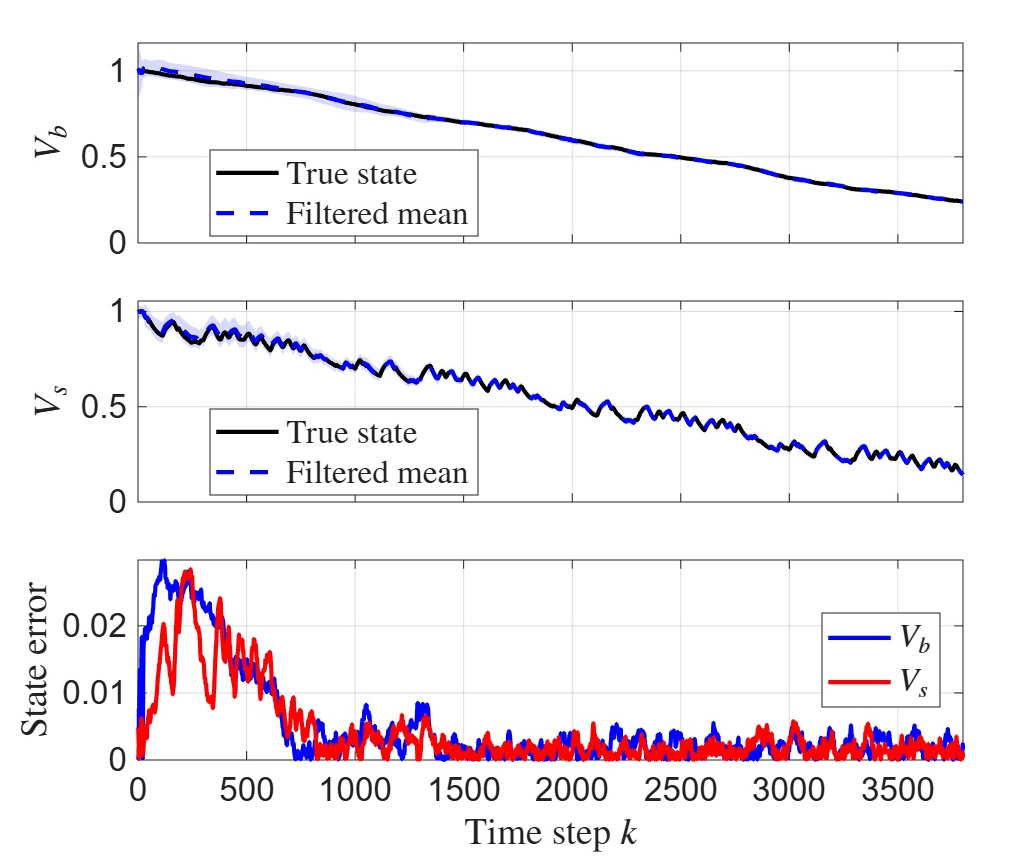}
\caption{Filtered estimates of the latent states $V_b$ and $V_s$ together with their estimation errors.}
\label{fig:states}
\end{figure}

Figure~\ref{fig:states} presents the filtered estimates of the latent states $V_b$ and $V_s$ together with their estimation errors. The filtered mean trajectories closely follow the true state evolution over the full simulation horizon. The estimation error is larger during the initial stage, particularly for $V_b$, due to the prior mismatch and the transient adaptation of the recursive estimator. As more measurements are incorporated, the errors decrease and remain small, showing that the particle-based state-inference step provides accurate recursive tracking of the latent NDC states.

\begin{figure}[htbp]
\vspace{2mm}
\centering
\includegraphics[width=1\linewidth]{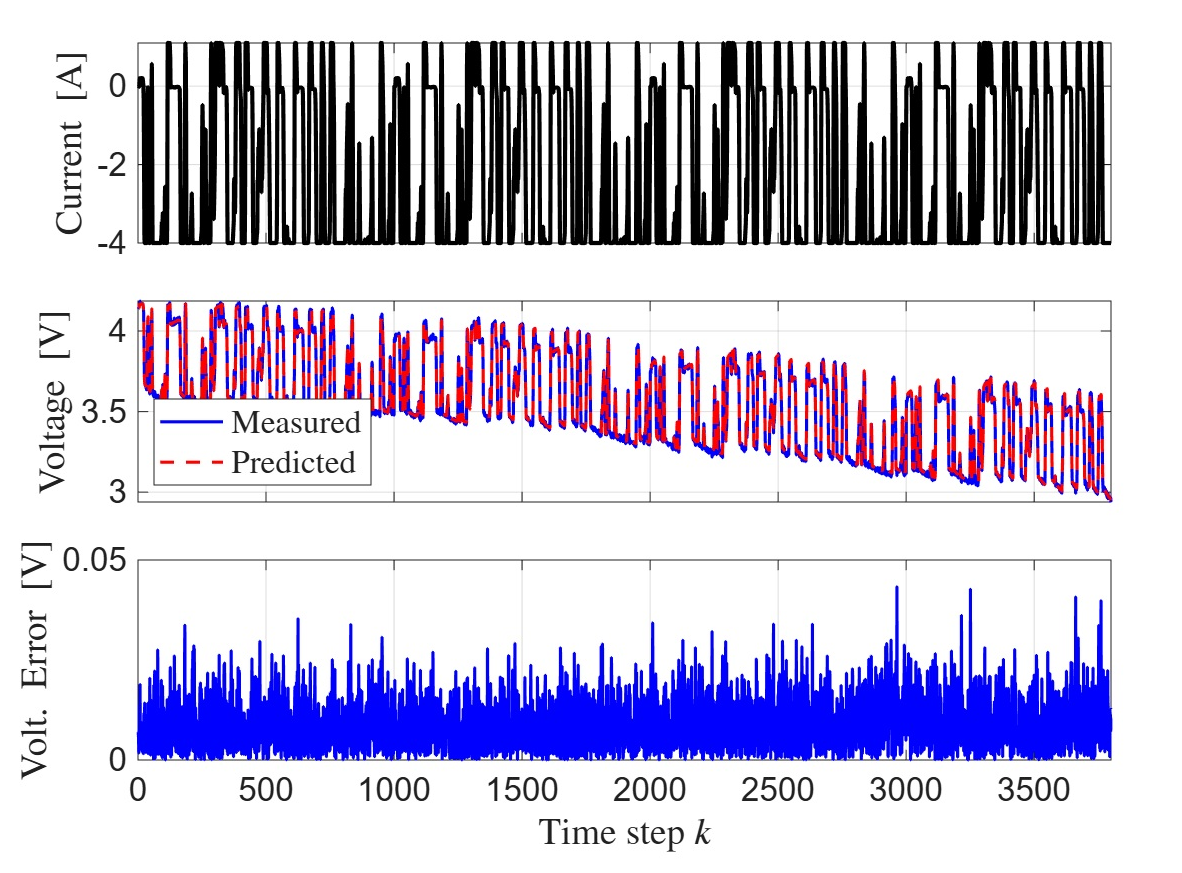}
\caption{Measured and reconstructed terminal voltage, together with the corresponding error, using the final SEnKI parameter estimate.}
\label{fig:voltage}
\end{figure}

Figure~\ref{fig:voltage} compares the measured terminal voltage with the
reconstructed voltage obtained using the final parameter estimate and shows
the corresponding error. The reconstructed voltage closely follows the measured trajectory over the full simulation horizon, and the reconstruction error remains small without noticeable systematic drift. Although local fluctuations appear in some operating intervals, the overall agreement confirms that the sequentially updated parameter ensemble captures the main input--output behavior of the NDC model.

\section{Conclusion}
\label{sec:conclusion}

Recursive parameter estimation for nonlinear stochastic SSMs remains a challenging problem despite its broad relevance to online model calibration and monitoring. A key difficulty arises from the coexistence of unknown parameters and latent states within nonlinear stochastic dynamics. To address this challenge, we have developed a new approach, termed SEnKI. The proposed method uses a parameter ensemble to approximate the posterior distribution of the unknown parameters and recursively updates the ensemble as new measurements arrive. In parallel, implicit particle filtering is employed to estimate the latent states conditioned on the latest parameter estimate. These two procedures alternate between parameter and state estimation, enabling recursive parameter learning while accounting for latent-state uncertainty. The ensemble-based representation provides a flexible means of capturing parameter uncertainty and facilitates accurate estimation in nonlinear systems. The proposed method has been evaluated on a modified UNGM, where it achieves higher estimation accuracy than the benchmark methods considered. It has further been applied to a lithium-ion battery model, with simulation results demonstrating its effectiveness in recursively identifying model parameters from measurement data.

\balance
\bibliographystyle{IEEEtran}
\bibliography{references}

\end{document}